\documentclass[12pt]{article}
\usepackage{graphicx}
\date{}
\def\be{\begin{equation}}
\def\ee{\end{equation}}
\def\bea{\begin{eqnarray}}
\def\eea{\end{eqnarray}}
\def\s{\sigma}
\def\al{\alpha}

\def\de{\delta}
\def\om{\omega}

\def\bc{\breve c}
\def\bs{\breve s}
\def\bC{\breve C}
\def\bS{\breve S}
\def\f{\varphi}

\title{UNSTABLE ROTATIONAL STATES\\
OF CLOSED STRING WITH MASSIVE POINTS
 }
\author{G.\,S. Sharov\\
{\small Tver state university}\\
{\small Tver, 170002, Sadovyj per. 35, Mathem. dep-t}}
\begin{document}
\maketitle
\begin{abstract}
For the closed string carrying 2 or 3 point-like masses the
stability problem for central and linear rotational states is
considered. This problem is important for applications of these
model to describing baryons, glueballs or other exotic hadrons.
The linear rotational state correspond to an uniform rotation of
the system with rectilinear string segments, connecting massive
points. The state is named ``central'' one, if there is a massive
point at the rotational center.

It is shown that the linear rotational states with 2 massive
points are stable with respect to small disturbances. But the
central rotational states with 3 masses are not stable, if the
central mass it less than energy of the string with other massive
points. This effect may change properties of excited hadron
states, in particular, increase their width.
\end{abstract}

\centerline {\bf 1. Introduction}
\medskip

Various string models of mesons, baryons and exotic hadrons
include a system of massive points connected by strings
\cite{Ch}\,--\,\cite{GlueY08}. In these models the massive points
describe quarks or constituent gluons and the Nambu-Goto string
simulates strong interaction between them and describes the QCD
confinement mechanism.

The mentioned models includes the string with massive ends as the
meson string model $q$-$\overline q$ \cite{Ch} or the
quark-diquark baryon model $q$-$qq$ \cite{Ko}; the linear string
baryon model $q$-$q$-$q$ \cite{lin,4B}; the  Y baryon model
${}^q_{\rule{0mm}{0.7em}\;\;\,q}\!\!\!\mbox{\textsf{Y}}^q_{\rule{0mm}{0.5em}}$
\cite{AY,PY}; the ``triangle'' (or $\Delta$) baryon configuration
 ${}^{\,}_q\!\!\stackrel{q}\triangle_q$ \cite{Tr},
and (generalizing the last model) the closed string with $n$
massive ends \cite{clbgl07,GlueY08}.

If we use a set of rotational states (planar uniform rotations of
a system), all the listed string hadron models generate linear or
quasilinear Regge trajectories $J\simeq\al_0+\al'E^2$, and may be
applied for describing excited hadron states with high angular
momenta $J$ are energies (masses) $E$
\cite{4B}\,--\,\cite{GlueY08}.

In these applications the problem of stability for rotational
states with respect to small disturbances is very important for
choosing the most adequate string model for baryons \cite{4B}, for
glueballs and other exotic hadrons
\cite{clbgl07}\,--\,\cite{KrivM5}. Another reason for interest to
this stability problem is the existence of quasirotational states
\cite{stab,qrottmf} in the linear vicinity of {\sl stable}
rotational states describing radial excitations or daughter Regge
trajectories for hadrons.

The stability problem for rotational states is solved for the
string with massive ends \cite{stab,qrottmf}, for the linear
string baryon model $q$-$q$-$q$ \cite{stlin}; and for the Y baryon
configuration \cite{Y02}. Analytical investigations and numerical
simulations of small disturbances demonstrated that rotational
states of the string with massive ends are stable
\cite{stab,qrottmf}, but they are unstable for string baryon
models $q$-$q$-$q$ and  Y \cite{stlin,Y02}. For the last two
models there are exponentially growing modes in spectra of small
disturbances.

In this paper the stability problem is solved for the certain
class of rotational states of the closed string carrying $n$
point-like masses. We consider the case $n=3$ and the rotational
states (named central states) with a massive point at the
rotational center. In the particular case, if the central mass
equals zero, this state is named linear rotational state with
$n=2$. The considered model describes baryons  \cite{Tr} or
glueballs \cite{clbgl07,GlueY08}.

Note that previously the stability problem for rotational states
was solved for the closed string with $n=1$ massive point
\cite{MilSh}.

\bigskip

\centerline {\bf 2. Dynamics and central rotational states}
\medskip

The classical dynamical equations for the closed string with
tension $\gamma$ carrying $n$ point-like masses $m_1$,
$m_2,\;\dots\,m_n$ result from the action \cite{Tr,clbgl07}
 $$
A=-\gamma\int\sqrt{(\dot X,X^{'})^2-\dot X^2 X^{'2}}\;d\tau d\s
-\sum\limits_{i=1}^n m_i\int\sqrt{\dot x_i^2(\tau)}\;d\tau,
 $$
 and have the form
 \be \frac{\partial^2X^\mu}{\partial\tau^2}-
\frac{\partial^2X^\mu}{\partial\s^2}=0
 \label{eq}\ee
 \be m_j\frac d{d\tau}\frac{\dot
x_j^\mu(\tau)}{\sqrt{\dot x_j^2(\tau)}}+\gamma
\Big[X^{'\!\mu}+\dot\s_j(\tau)\dot
X^\mu\Big]\Big|_{\s=\s_j-0}-\gamma\Big[X^{'\!\mu}+\dot\s_j(\tau)\dot
X^\mu\Big]\Big|_{\s=\s_j+0}=0,
 \label{qqi}\ee
 \be m_n\frac d{d\tau}\frac{\dot x_0^\mu(\tau)}{\sqrt{\dot
x_0^2(\tau)}}+\gamma
\Big[X^{'\!\mu}\big(\tau^*(\tau),2\pi\big)-X^{'\!\mu}(\tau,0)\Big]=0.
\label{qq0}\ee
 if the orthonormality conditions on the world surface $X^\mu(\tau,\s)$
 \be
(\partial_\tau X\pm\partial_\s X)^2=0,
 \label{ort}\ee
 and the conditions
 \be \s_0(\tau)=0,\qquad \s_n(\tau)=2\pi,
\label{ends}\ee
 are fulfilled  \cite{Tr,GlueY08}.
 Here $\dot
X^\mu\equiv\partial_\tau X^\mu$, $X^{'\!\mu}\equiv\partial_\s
X^\mu$, Scalar product in Minkowski space $R^{1,3}$ is
$(a,b)=\eta_{\mu\nu}a^\mu b^\nu$,
$\eta_{\mu\nu}={}$diag$(1,-1,-1,-1)$; speed of light $c=1$, the
parameter $\s$ varies in the limits $\s_0\le\s\le\s_n$, that is
$\s\in[0,2\pi]$, equations $\s=\s_j(\tau)$ and
 $$x^\mu=x_j^\mu(\tau)\equiv X^\mu\big(\tau,\s_j(\tau)\big),\qquad j=0,1,\dots,n$$
determine world lines of the massive points, for the case $j=0$
and $j=n$ they describe the same trajectory of the $n$-th point,
and their equality forms the closure condition
 \be
X^\mu(\tau^*,2\pi)=X^\mu(\tau,0)
 \label{clos}\ee
 on the tube-like world surface \cite{Tr,MilSh}. These equations may
contain two different parameters $\tau$ and $\tau^*$, connected
via the relation $\tau^*=\tau^*(\tau)$. This relation should be
included in the closure condition (\ref{clos}) of the world
surface.

 Conditions (\ref{ort}), (\ref{ends}) always may be
fixed without loss of generality, if we choose the relevant
coordinates $\tau$, $\s$ \cite{Tr}.

Eqs.~(\ref{qqi}), (\ref{qq0}) are equations of motion for the
massive points resulting from the action. They may be interpreted
as boundary conditions for Eq.~(\ref{eq}).

The system of equations (\ref{eq})\,--\,(\ref{clos}) describe
dynamics of the closed string carrying $n$ point-like masses
without loss of generality. One also should add that a tube-like
world surface of the closed string is continuous one, but its
derivatives may have discontinuities at world lines of the massive
points (except for derivatives along these lines) \cite{Tr}. These
discontinuities are taken into account in Eqs.~(\ref{qqi}),
(\ref{qq0}).

Exact solutions of the system (\ref{eq})\,--\,(\ref{clos})
describing rotational states of the closed string with masses we
obtained and classified in Refs.~\cite{clbgl07,GlueY08}. These
states are divided into three classes: hypocycloidal, linear and
central rotational states. In hypocycloidal states the rotating
string is composed of segments of a hypocycloid. For linear states
these segments are rectilinear and all masses $m_j$ move at
nonzero velocities $v_j$. Central states are states with a massive
point (or some of them) placed at the rotational center.

The string world surface for the central or linear states can be
presented as \cite{clbgl07}
 \be
X^\mu(\tau,\s)=x_0^\mu+e_0^\mu a_0\tau+ u(\s)\cdot e^\mu(\om\tau).
 \label{Xlin}\ee
 Here
 \be
 e^\mu(\om\tau)=e^\mu_1\cos\om\tau+e^\mu_2\sin\om\tau,
 \label{e}\ee
  the unit vectors
 $$ e^\mu_0,\;e^\mu_1,\;e^\mu_2,\;e^\mu_3,$$
 associated with coordinates $x^\mu$, form the orthonormal basis in $R^{1,3}$.
In the closure  condition (\ref{clos}) $\tau^*=\tau$. For the
solution (\ref{Xlin}) of Eq.~(\ref{eq}) the following values are
the constants:
 \be
  \s_j(\tau)=\s_j={}\mbox{const},\qquad j=1,2,
 \label{si12}\ee
 \be
Q_j=\frac\gamma{m_j}\sqrt{\dot x_j^2(\tau)},\qquad
 j=1,2,3.
 \label{Qi}\ee
 The function
 \be
u(\s)=\frac{a_0}\om\cdot\left\{\begin{array}{ll}\sin\om\s,&
\s\in[0,\s_1],\\
2\bs_1\bc_1\cos\om\s+(\bs_1^2-\bc_1^2)\sin\om\s,&\s\in[\s_1,\s_2],\\
-\bS\cos\om\s+\bC\sin\om\s,&\s\in[\s_2,2\pi]
\end{array}\right.
 \label{u}\ee
 is determined from Eq.~(\ref{eq}), continuity of the
function $X^\mu(\tau,\s)$ on the lines $\s=\s_j$ and conditions
(\ref{qqi})\,--\,(\ref{clos}) \cite{clbgl07}. Here and below we
use the following notations for constants:
 $$\begin{array}{c}
\bc_1=\cos\om\s_1,\qquad \bs_1=\sin\om\s_1,\qquad
\bC=\cos2\pi\om,\qquad \bS=\sin2\pi\om,\\
\bc_3=\cos\om(2\pi-\s_2),\qquad \bs_3=\sin\om(2\pi-\s_2),\qquad
\bC_2=\cos\om\s_2,\qquad \bS_2=\sin\om\s_2.\rule{0mm}{1.2em}
 \end{array}$$

Expression (\ref{Xlin}) describes a rotation of two coinciding
string segments, connecting two points with masses $m_1$, $m_2$.
These points move along circles at velocities $v_1$ and $v_2$, the
system rotates in the plane $e_1,\,e_2$ at angular velocity
$\Omega=\om/a_0$. The point with mass $m_3$ is at rest at the
rotational center ($v_3=0$). Values of parameters $\s_j$, $Q_j$,
$m_j$, $v_j$, $a_0$ are related by the following equations,
resulting from Eqs.~(\ref{qqi})\,--\,(\ref{clos}) \cite{clbgl07}:
 \be
\s_2-\s_1=\pi,
 \label{s21}\ee
 \be
  v_1=\bs_1=\sin\om\s_1,\qquad v_2=\bs_3=\sin\om(\pi-\s_1),
 \label{v1v2}\ee
 \be
2\frac{Q_1}\om=\frac{\bs_1}{\bc_1}=\frac{v_1}{\sqrt{1-v_1^2}},\qquad
2\frac{Q_2}\om=\frac{\bs_3}{\bc_3}=\frac{v_2}{\sqrt{1-v_2^2}},
 \label{h1h2}\ee
 \be a_0
=\frac{m_1Q_1}{\gamma\sqrt{1-v_1^2}}=\frac{m_2Q_2}{\gamma\sqrt{1-v_2^2}}=
\frac{m_3Q_3}{\gamma},
 \label{a0v}\ee
 \be
\frac{m_1v_1}{1-v_1^2}=\frac{m_2v_2}{1-v_2^2}.
 \label{mv12}\ee

If the values $v_1$ and $v_2$ are given, one can calculate from
Eqs.~(\ref{v1v2}) the following values:
 \be
\om=\frac{(-1)^{k_1}\arcsin v_1+(-1)^{k_2}\arcsin
v_2}\pi+2(k_1+k_2),\quad \s_1=\frac{(-1)^{k_1}\arcsin v_1+2\pi
k_1}\om.
 \label{oms1}\ee
  Here $k_1$ and $k_2$ are arbitrary integers resulting in the inequalities
$0<\s_1<\pi$, $\bs_1>0$, $\bs_3>0$. The simplest case $k_1=k_2=0$
corresponds to 2 coinciding rectilinear string segments,
connecting the points $m_1$, $m_2$. For other permissible values
$k_1$ and $k_2$ these segments are kinked curves, folded on one
(rotating) straight line. The fold points move along circles at
the speed of light.

\bigskip

\centerline {\bf 3. Stability problem for central rotational
states}

\medskip

Possible applications of solutions (\ref {Xlin}) in hadron
spectroscopy essentially depend on stability or instability of
these states with respect to small disturbances. In this section
we study spectrum of these disturbances for the central rotational
states.

This problem has been recently solved for the closed string with
$n=1$ massive point in Ref.~\cite{clstab05}. In
Ref.~\cite{clbgl07} central rotational states with $n=3$ are
considered, but some simplifying assumptions in this paper
appeared to be incorrect, and the results need refinement.

To solve the stability problem for the central rotational states
(\ref{Xlin}) we consider the general solution of Eq.~(\ref{eq})
for the string with $3$ masses
 \be
X^\mu (\tau,\s)=\frac{1}{2}[\Psi^\mu_{j+}(\tau +\s)+\Psi
^\mu_{j-}(\tau-\s)],\qquad\s\in[\s_{j-1},\s_j],\quad j=1,2,3.
 \label{gensol}\ee
 Here the functions $\Psi^\mu_{j\pm}(\tau\pm\s)$ are smooth, the world surface
(\ref{gensol}) is smooth between world lines of massive points.

We denote $\Psi^{(r)\mu}_{j\pm}$ the functions in the expression
(\ref {gensol}) for the rotational states (\ref{Xlin}). Their
derivatives
 $\dot\Psi^{(r)\mu}_{j\pm}\equiv\frac d{d\tau}\Psi^{(r)\mu}_{j\pm}$
in accordance with Eq.~(\ref{u}) are
 \be \begin{array}{l}
 \dot\Psi^{(r)\mu}_{1\pm}(\tau)=a_0\Big[e_0^\mu\pm
e^\mu(\om\tau)\Big],\\
 \dot\Psi^{(r)\mu}_{2\pm}(\tau)=a_0\Big[e_0^\mu+2v_1\bc_1\acute
e^\mu(\om\tau)\pm(2v_1^2-1)\,e^\mu(\om\tau)\Big],\rule{0mm}{6.5mm}\\
 \dot\Psi^{(r)\mu}_{3\pm}(\tau)=a_0\Big[e_0^\mu-\breve S\acute
e^\mu(\om\tau)\pm\breve C e^\mu(\om\tau)\Big],\rule{0mm}{6.5mm}
\end{array}
 \label{Psic}\ee
 Here the rotating vector
 $$\acute e^\mu(\om\tau)=-e^\mu_1\sin\om\tau+e^\mu_2\cos\om\tau,$$
 is orthogonal to the vector $e^\mu(\om\tau)$ (\ref{e}).

To describe any small disturbances of the rotational motion, that
is motions close to states (\ref{Xlin}) we consider vector
functions $\dot\Psi^{\mu}_{j\pm}$ close to
$\dot\Psi^{(r)\mu}_{j\pm}$ in the form
 \be
\dot\Psi^\mu_{j\pm}(\tau)=\dot\Psi^{(r)\mu}_{j\pm}(\tau)
+\f_{j\pm}^\mu(\tau).\label{Psi+f}\ee

The disturbance $\f_{j\pm}^\mu(\tau)$ is supposed to be small, so
we omit squares of $\f_{j\pm}$ when we substitute the expression
(\ref{Psi+f}) into dynamical equations (\ref{qqi}), (\ref{qq0})
and (\ref{clos}). In other words, we work in the first linear
vicinity of the states  (\ref{Xlin}). Both functions
$\Psi^\mu_{j\pm}$ and $\Psi^{(r)\mu}_{j\pm}$ in expression
(\ref{Psi+f}) must satisfy the condition
 $${\dot\Psi_{j+}\!\!\!\!}^2={\dot\Psi_{j-}\!\!\!\!}^2=0,$$
 resulting from Eq.~(\ref{ort}), hence in the first order approximation in $\f_{j\pm}$
the following scalar product equals zero:
 \be
\big(\dot\Psi^{(r)}_{j\pm},\f_{j\pm}\big)=0.
 \label{Psif}\ee

For the disturbed motions the equalities $\tau^*=\tau$ and
$\s_j(\tau)=\s_j={}$const (\ref{si12}), generally speaking, is not
carried out and should be replaced with the equalities
 \be
\s_1(\tau)=\s_1+\de_1(\tau),\qquad\s_2(\tau)=\s_2+\de_2(\tau),
\qquad\tau^*=\tau+\de(\tau),
 \label{taudel}\ee
 where $\de_j(\tau)$ and $\de(\tau)$ are small disturbances.

Expression (\ref{Psi+f}) together with Eq.~(\ref{gensol}) is the
solution of the  string motion equation (\ref {eq}). Therefore we
can obtain equations of evolution for small disturbances
$\f_{j\pm}^\mu(\tau)$, substituting expressions (\ref{Psi+f}) and
(\ref{taudel}) with Eq.~(\ref{Psic}) into other equations of
motion (\ref{qqi}), (\ref{qq0}), the closure condition
(\ref{clos}) and the continuity condition
 \be
X^\mu\big(\tau,\s_j(\tau)-0\big)=X^\mu\big(\tau,\s_j(\tau)+0\big),\qquad
i=1,2
 \label{cont}\ee
 in linear  approximation.
 We are to take into account nonlinear factors
$\Big\{\big[\frac
d{d\tau}X\big(\tau,\s_j(\tau)\big)\big]^2\Big\}^{-1/2}$ and
contributions from the disturbed arguments $\tau^*$ and
$\s_j(\tau)$ (\ref{taudel}), for example:
$$
\dot\Psi^{(r)\mu}_{3\pm}(\tau^*\pm2\pi)\simeq
\dot\Psi^{(r)\mu}_{3\pm}(\tau\pm2\pi)+\de(\tau)\,
\ddot\Psi^{(r)\mu}_{3\pm}(\tau\pm2\pi).
$$

This substitution for the central rotational state (\ref{Xlin})
with $n=3$ and vector-functions $\dot\Psi^{(r)\mu}_{j\pm}$
(\ref{Psic}) after simplifying results in the following system of
6 vector equations in linear (with respect to $\f_{j\pm}^{\mu}$,
$\de_j$ and $\de$) approximation:
 \be
\begin{array}{c}
\f_{1+}^\mu(+_1)+\f_{1-}^\mu(-_1)-\f_{2+}^\mu(+_1)-\f_{2-}^\mu(-_1)+4\bc_1a_0
\big[e^\mu(\om\tau)\,\dot\de_1(\tau)+\om\acute
e^\mu(\om\tau)\,\de_1\big]=0,\\
\f_{2+}^\mu(+_2)+\f_{2-}^\mu(-_2)-\f_{3+}^\mu(+_2)-\f_{3-}^\mu(-_2)-4\bc_3a_0
\big[e^\mu(\om\tau)\,\dot\de_2(\tau)+\om\acute
e^\mu(\om\tau)\,\de_2\big]=0,\rule{0mm}{5.5mm}\\
\f_{3+}^\mu(+)+\f_{3-}^\mu(-)-\f_{1+}^\mu(\tau)-\f_{1-}^\mu(\tau)+2a_0e_0^\mu
\dot\de(\tau)=0,\rule{0mm}{5.5mm}\\
\frac d{d\tau}\Big\{\f_{1+}^\mu(+_1)+\f_{1-}^\mu(-_1)+2\bc_1a_0
(e^\mu\dot\de_1+\om\acute e^\mu\de_1)+G_1(e_0^\mu+v_1\acute
e^\mu)\Big\}+{}
\rule{0mm}{6.8mm}\\
\qquad{}+Q_1\Big[\f_{1+}^\mu(+_1)-\f_{1-}^\mu(-_1)-\f_{2+}^\mu(+_1)+\f_{2-}^\mu(-_1)\Big]=0.
\rule{0mm}{6.0mm}\\
\frac d{d\tau}\Big\{\f_{2+}^\mu(+_2)+\f_{2-}^\mu(-_2)-2\bc_3a_0
(e^\mu\dot\de_2+\om\acute e^\mu\de_2)+G_2(e_0^\mu-v_2\acute
e^\mu)\Big\}+{}
\rule{0mm}{6.80mm}\\
\qquad{}+Q_2\Big[\f_{2+}^\mu(+_2)-\f_{2-}^\mu(-_2)-\f_{3+}^\mu(+_2)+\f_{3-}^\mu(-_2)\Big]=0.
\rule{0mm}{6.0mm}\\
\!\!\!\frac
d{d\tau}\Big\{\f_{1+}^\mu+\f_{1-}^\mu+(\f_{1+}-\f_{1-})\,e_0^\mu\Big\}+
Q_3\Big[\f_{3+}^\mu(+)-\f_{3-}^\mu(-)-\f_{1+}^\mu+\f_{1-}^\mu+2\om
a_0\acute e^\mu\de\Big]=0. \rule{0mm}{6.5mm}\!\!\!
\end{array} \label{sysf}\ee
 In the last three equations arguments $(\tau)$ for $\f_{1\pm}^\mu$,
 $\de$, $\de_j$ and $(\om\tau)$ for $e^\mu$,
$\acute e^\mu$ are omitted; we use the following notations for
arguments
 $$
(\pm_1)\equiv(\tau\pm\s_1),\qquad
(\pm_2)\equiv(\tau\pm\s_2),\qquad(\pm)\equiv(\tau\pm2\pi),
 $$
 for the scalar products
 \be
  \f_{j\pm}^0\equiv( e_0,\f_{j\pm}),\qquad
\f_{j\pm}^3\equiv( e_3,\f_{j\pm}),\qquad \f_{j\pm}\equiv (
e,\f_{j\pm}), \qquad \acute\f_{j\pm}\equiv(\acute e,\f_{j\pm})
\label{fiscal}\ee
 and
 $$
\begin{array}{l}
G_1=\f_{1+}(+_1)-\f_{1-}(-_1)-v_1\bc_1^{-1}\Big[\acute\f_{1+}(+_1)+\acute\f_{1-}(-_1)-
2\om a_0\de_1\Big],\\
G_2=\bc_3^{-1}\Big\{\breve C_2\big[\f_{2-}(-_2)-\f_{2+}(+_2)\big]+
\breve S_2\big[\acute\f_{2+}(+_2)+\acute\f_{2-}(-_2)\big]+ 2\om
v_2a_0\de_2\Big\}.\rule{0mm}{6.0mm}
\end{array}
 $$

The first two equations (\ref{sysf}) results from
Eqs.~(\ref{cont}), the third ---  from Eq.~(\ref{clos}), other
ones are consequence of Eqs.~(\ref{qqi}) and (\ref{qq0}).
Equations (\ref{sysf}) are simplified with using
Eqs.~(\ref{s21})\,--\,(\ref{oms1}), (\ref{Psic}) and equalities
(\ref{Psif}), resulting in the following relations for projections
(\ref{fiscal}) of disturbances:
 \be
\f_{1\pm}^0(\tau)=\mp\f_{1\pm}(\tau),\qquad
\f_{2\pm}^0=\mp(2v_1^2-1)\,\f_{2\pm}-2v_1\bc_1\acute\f_{2\pm},\qquad
\f_{3\pm}^0=\breve S\acute\f_{3\pm}\mp\breve C\f_{3\pm}.
 \label{eqscal}\ee

The linearized system of equations (\ref{sysf}), (\ref{eqscal})
describes evolution of small disturbances of the considered
central rotational state (\ref{Xlin}), (\ref{Psic}).

Note that scalar products of Eqs.~(\ref{sysf}) onto the vector
$e_3$ (orthogonal to the rotational plane $e_1$,\,$e_2$) form the
closed subsystem from 6 equations with respect to 6 functions
(\ref{fiscal}) $\f_{j\pm}^3$:
 \be
\begin {array}{c}
\f_{j+}^3(+_j)+\f_{j-}^3(-_j)=\f_{j^*+}^3(+_j)+\f_{j^*-}^3(-_j),\\
\f_{3+}^3(+)+\f_{3-}^3(-)=\f_{1+}^3(\tau)+\f_{1-}^3(\tau),\\
\dot\f_{j+}^3(+_j)+\dot\f_{j-}^3(-_j)+Q_j\Big[
\f_{j+}^3(+_j)-\f_{j-}^3(-_j)-\f_{j^*+}^3(+_j)+\f_{j^*-}^3(-_j)\Big]=0,
\rule{0mm}{5mm}\\
 \dot\f_{1+}^3(\tau)+\dot\f_{1-}^3(\tau)+Q_3\Big[
\f_{3+}^3(+)-\f_{3-}^3(-)-\f_{1+}^3(\tau)+\f_{1-}^3(\tau)\Big]=0.
\rule{0mm}{5mm}\end{array}
 \label{sysf3}\ee
 Here $j=1,2$, $j^*\equiv j+1$. This system is homogeneous system
with deviating arguments.

We search solutions of this system in the form of harmonics
 \be
\f_{j\pm}^3=B_{j\pm}^3 \exp(-i\xi\tau).
 \label{fexp3} \ee

This substitution results in the linear homogeneous system of 6
algebraic equations with respect to 6 amplitudes $B_{j\pm}^3$. The
system has nontrivial solutions if and only if its determinant
 $$ \left|
\begin{array}{cccccc}
E_{1+} & E_{1-}& -E_{1+} & -E_{1-}& 0 & 0 \\
 0 & 0 & E_{2+} & E_{2-}& -E_{2+} & -E_{2-}\\
-1 & -1 & 0 & 0 & E_{3+} & E_{3-}\\
(i\xi-Q_1)\,E_{1+}& (i\xi+Q_1)\,E_{1-}&Q_1E_{1+}& -Q_1E_{1-} & 0 & 0 \\
 0 & 0 & (i\xi-Q_2)\,E_{2+}& (i\xi+Q_2)\,E_{2-}& Q_2E_{2+} &-Q_2E_{2-} \\
-i\xi-Q_3 &-i\xi+Q_3 & 0 & 0 & Q_3E_{3+} &-Q_3E_{3-}
\end{array}\right|=0
 $$
 equals zero. Here $E_{j\pm}=\exp(\mp i\xi\s_j)$. This equation is reduced to
 the form
$$
2(\cos2\pi\xi-1)-\xi\big(Q_1^{-1}+Q_2^{-1}+Q_3^{-1}\big)\sin2\pi\xi+{}
 $$
 \be
{}+\xi^2\left(\frac{\sin^2\pi\xi}{Q_1Q_2}+ \frac{\tilde s_3
\sin\s_2\xi}{Q_2Q_3}+ \frac{\tilde
s_{23}\sin\s_1\xi}{Q_1Q_3}\right)- \frac{\xi^3\tilde
s_3\sin\s_1\xi\cdot\sin\pi\xi}{Q_1Q_2Q_3}=0,
 \label{tom3}\ee
 where $\tilde s_3=\sin(\pi-\s_1)\xi$, $\tilde s_{23}=\sin(2\pi-\s_1)\xi$.

This equation describes the spectrum of frequencies $\xi$ for
transversal (with respect to the $e_1,\,e_2$ plane) small
fluctuations of the string for the considered rotational state. If
equation (\ref{tom3}) has a complex root $\xi=\xi_1+i\xi_2$ with
positive imaginary part $\xi_2$, the amplitude of the
correspondent disturbance will grow exponentially:
 $$
 \f_{k}^3=B_{k}^3\exp(-i\xi_1\tau)\cdot\exp(\xi_2\tau).
 $$
 In this case the considered rotational state is unstable one \cite{stab,Y02}.

\begin{figure}[ht]
\includegraphics[scale=0.85,trim=30 25 10 105]{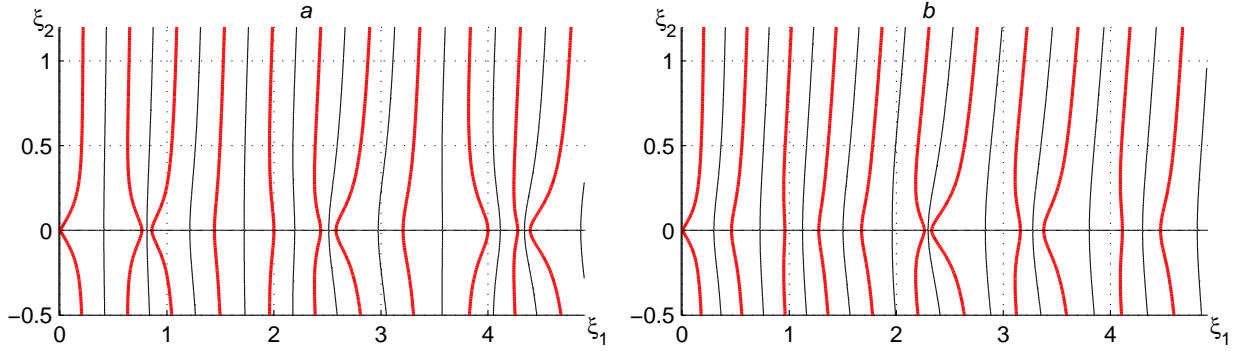}
\caption{Zero level lines for real part (thick) and imaginary part
(thin) of Eq.~(\ref{tom3}); {\it a)} $Q_1=Q_2=Q_3=1$, {\it b)}
$Q_1=1$, $Q_2=0{.}4$, $Q_3=0{.}1$}
\end{figure}

Analysis of the real and imaginary parts of Eq.~(\ref{tom3}) is
presented in Fig.~1, where the thick and thin lines are zero level
lines correspondingly for real and imaginary part of the function
$f(\xi)=f(\xi_1+i\xi_2)$ in Eq.~(\ref{tom3}) for given values
$Q_j$. Roots of this equation are shown as cross points of a thick
line with a thin line. If the values (\ref{Qi}) $Q_j$ are given,
one can determine values $\om$, $\s_1$, $v_j$ from
Eqs.~(\ref{s21})\,--\,(\ref{oms1}). For example, $\om$ is one of
the roots of the equation
 \be
\tan\pi\om=\frac{2\om(Q_1+Q_2)}{\om^2-4Q_1Q_2},
 \label{omQ12}\ee
 resulting from Eqs.~(\ref{h1h2}). For the state in  Fig.~1{\it a}
the root of this equation $\om\simeq0{.}766898$ is chosen. It
corresponds to the mass relation $m_1:m_2:m_3=1:1:2{.}79$; for the
state in Fig.~1{\it b} $\om\simeq2{.}331$,
$m_1:m_2:m_3=1:3{.}12:13{.}18$. The frequency $\om$ for the state
in Fig.~1{\it a} is the minimal positive root of
Eq.~(\ref{omQ12}), corresponding to $k_1=k_2=0$ in
Eqs.~(\ref{oms1}) (the simplest rectilinear string segments,
connecting the massive points). For the state in Fig.~1{\it b} the
values in Eqs.~(\ref{oms1}) $k_1=0$, $k_2=1$, and the string
segments are kinked lines.

The analysis of Eq.~(\ref{tom3}) for various values $Q_j$ and
$\om$ demonstrates that all its roots are real
numbers, therefore amplitudes of such fluctuations do not grow
with growth of time $t$.

\bigskip

\centerline {\bf 4. Disturbances in the rotational plane}
\medskip

One can not solve the stability problem for the cental rotational
states (\ref{Xlin}) only on the base of the above studied behavior
of disturbances $\f_{j\pm}^3$ (\ref{fexp3}) (orthogonal to the
rotational plane). We are to consider small disturbances
concerning to the $e_1,\,e_2$ plane. Projections (scalar products)
of equations (\ref{sysf}) onto 3 vectors $e_0$, $e(\tau)$, $\acute
e(\tau)$ with using relations
 $$
e^\mu(\tau)=\bc_1e^\mu(\pm_1)\mp\bs_1\acute e^\mu(\pm_1)
=\bC_2e^\mu(\pm_2)\mp\bS_2\acute e^\mu(\pm_2)= \bC e^\mu(\pm)\mp
\bS\acute e^\mu(\pm)
 $$
 and their analogs for $\acute e^\mu$ (for multiplying by
vector-functions $\f^\mu_{j\pm}$ with different arguments) result
in the system of 18 equations. Three of them are linear
combinations of the other ones. The rest equations form the system
of 15 differential equations with deviating arguments with respect
to 15 unknown functions of $\tau$: $\f_{j\pm}$, $\acute\f_{j\pm}$
($j=1,2,3$), $\de_1$, $\de_2$, $\de$. The functions $\f_{j\pm}^0$
are excluded via Eqs.~(\ref{eqscal}). Here we present 8 equations
from this system, resulting from the 1-st, 3-rd and 6-th equations
(\ref{sysf}):
$$
\begin {array}{c}
\f_{1+}(+_1)-\f_{1-}(-_1)+(\bc_1^2-\bs_1^2)\big[\f_{2+}(+_1)-\f_{2-}(-_1)\big]-
2\bs_1\bc_1\big[\acute\f_{2+}(+_1)+\acute\f_{2-}(-_1)\big]=0,
\rule{0mm}{1.2em}\\
\bc_1\big[\f_{1+}(+_1)+\f_{1-}(-_1)-\f_{2+}(+_1)-\f_{2-}(-_1)-4a_0\dot\de_1(\tau)\big]+{}
\qquad\qquad
\rule{0mm}{1.2em}\\
\qquad\qquad\qquad{}+
\bs_1\big[\acute\f_{1-}(-_1)-\acute\f_{1+}(+_1)+\acute\f_{2+}(+_1)-\acute\f_{2-}(-_1)\big]=0,\\
\bs_1\big[\f_{1+}(+_1)-\f_{1-}(-_1)-\f_{2+}(+_1)+\f_{2-}(-_1)\big]+{}\qquad\qquad\qquad\qquad
\rule{0mm}{1.2em}\\
\qquad\qquad{}+
\bc_1\big[\acute\f_{1+}(+_1)+\acute\f_{1-}(-_1)-\acute\f_{2+}(+_1)-
\acute\f_{2-}(-_1)-4\om
a_0\de_1(\tau)\big]=0,\\
\f_{1+}(\tau)+\f_{1-}(\tau)-\bC\f_{3+}(+)-\bC\f_{3-}(-)+\bS\acute\f_{3+}(+)-\bS\acute\f_{3-}(-)
=0,\rule{0mm}{1.2em}\\
\acute\f_{1+}(\tau)+\acute\f_{1-}(\tau)-\bS\f_{3+}(+)+\bS\f_{3-}(-)-\bC\acute\f_{3+}(+)-
\bC\acute\f_{3-}(-)=0,\rule{0mm}{1.2em}\\
\f_{1+}(\tau)-\f_{1-}(\tau)-\bC\f_{3+}(+)+\bC\f_{3-}(-)+\bS\acute\f_{3+}(+)+\bS\acute\f_{3-}(-)
+2a_0\dot\de(\tau)=0,\rule{0mm}{1.2em}\\
\dot\f_{1+}+\dot\f_{1-}-\om(\acute\f_{1+}+\acute\f_{1-})=Q_3\big[\f_{1+}-\f_{1-}
-\bC(\f_{3+}-\f_{3-})+\bS(\acute\f_{3+}+\acute\f_{3-})\big],
\rule{0mm}{1.2em}\\
\dot{\acute\f}_{1+}+\dot{\acute\f}_{1-}+\om(\f_{1+}+\f_{1-})=
Q_3\big[\acute\f_{1+}-\acute\f_{1-}
-\bS(\f_{3+}+\f_{3-})-\bC(\acute\f_{3+}-\acute\f_{3-})+2\om
a_0\de\big]. \rule{0mm}{1.2em} \end{array}
$$
 In the last equations the arguments $(\tau)$ for
$\f_{1\pm}$, $\acute\f_{1\pm}$, $\de$, and $(\pm)$ for
$\f_{3\pm}$, $\acute\f_{3\pm}$ are omitted.

When we search solutions of this system in the form of harmonics
(\ref{fexp3})
 \be
\f_{j\pm}=B_{j\pm} e^{-i\xi\tau},\qquad \acute \f_{j\pm}=\acute
B_{j\pm} e^{-i\xi\tau}, \qquad 2a_0\de_j=\Delta_j e^{-i\xi\tau},
\qquad 2a_0\de=\Delta e^{-i\xi\tau},
 \label{fexp} \ee
 we obtain the homogeneous system of 15 algebraic equations with
respect to 15 amplitudes $B_{j\pm}$, $\acute B_{j\pm}$,
$\Delta_1$, $\Delta_2$, $\Delta$. The mentioned above condition of
existence of nontrivial solutions for this system is vanishing the
corresponding determinant.

This cumbersome determinant was calculated at the first stage in
the particular symmetric case of equal masses
 \be
m_1=m_2,\qquad\s_1=\frac\pi2.
 \label{mm} \ee
 with using symbolic calculations in the package MATLAB and numerical
verification.

In the case (\ref{mm}), resulting in equalities $\bs_1=\bs_3$,
$\bc_1=\bc_3$, $v_1=v_2$, $Q_1=Q_2$, the obtained equation of the
spectrum for small disturbances is equivalent to the set of
equations (is factorized):
  \be
\Big(\xi\tan\frac{\pi\xi}2-\om\tan\frac{\pi\om}2\Big)
\Big(\bc_1^2\xi^2-2\bs_1\bc_1\om\xi\cot\frac{\pi\xi}2-Z_1\om^2\Big)=0,
 \label{frmm1}\ee
 $$
\tilde c\,\xi(\xi^2-\om^2)\big[\tilde s\bc_1^3\xi^3-3\tilde
c\bs_1\bc_1^2\om\xi^2-\tilde s(1+3\bs_1^2)\,\bc_1\om^2\xi+\tilde
c\bs_1Z_1\om^3\big]=(\tilde c_1\bc_1^2\xi^2+2\tilde
s_1\bs_1\bc_1\om\xi-\tilde c_1Z_1\om^2)\times
 $$
  \be 
 \times4Q_3\Big[ 
 \bc_1\xi^3\cos\frac{3\pi\xi}2+2(\tilde c\tilde
s_1\bs_1\om+\tilde s\tilde c_1\bc_1Q_3)\,\xi^2+(\tilde c\tilde c_1
\bc_1\om+2\tilde s\tilde s_1\bs_1Q_3)\,\om\xi+\tilde c\tilde
s_1\bs_1\om^3\Big].
 \label{frmm2}\ee
 Here
 $$Z_1=1+\bs_1^2,\quad\tilde c=\cos\pi\xi,\quad\tilde s=\sin\pi\xi,\quad\tilde
 c_1=\cos\s_1\xi,\quad\tilde s_1=\sin\s_1\xi.$$

Analysis of equation (\ref{frmm1}) (decomposing into two factors)
for complex $\xi=\xi_1+i\xi_2$ shows, that for all values $\om$
and $Q_1=\frac12\om\bs_1/\bc_1$ all roots of this equation are
real numbers and form a countable set. Their behavior is similar
to that for roots of Eq.~(\ref{tom3}).

But roots of equation (\ref{frmm2}) have other properties. These
roots are shown in Fig.~2 as cross points of thick and thin lines
for two types of rotational states: for $Q_1=Q_2=1/4$, $\om=1/2$,
$v_1=v_2\simeq0{.}707$ on the left and $Q_1=Q_2=1$,
$\om\simeq0{.}766898$, $v_j\simeq0{.}934$ on the right for various
values $Q_3$ and corresponding $m_3=\gamma a_0/Q_3$ (\ref{a0v}).

\begin{figure}[th]
\includegraphics[scale=0.83,trim=25 8 10 3]{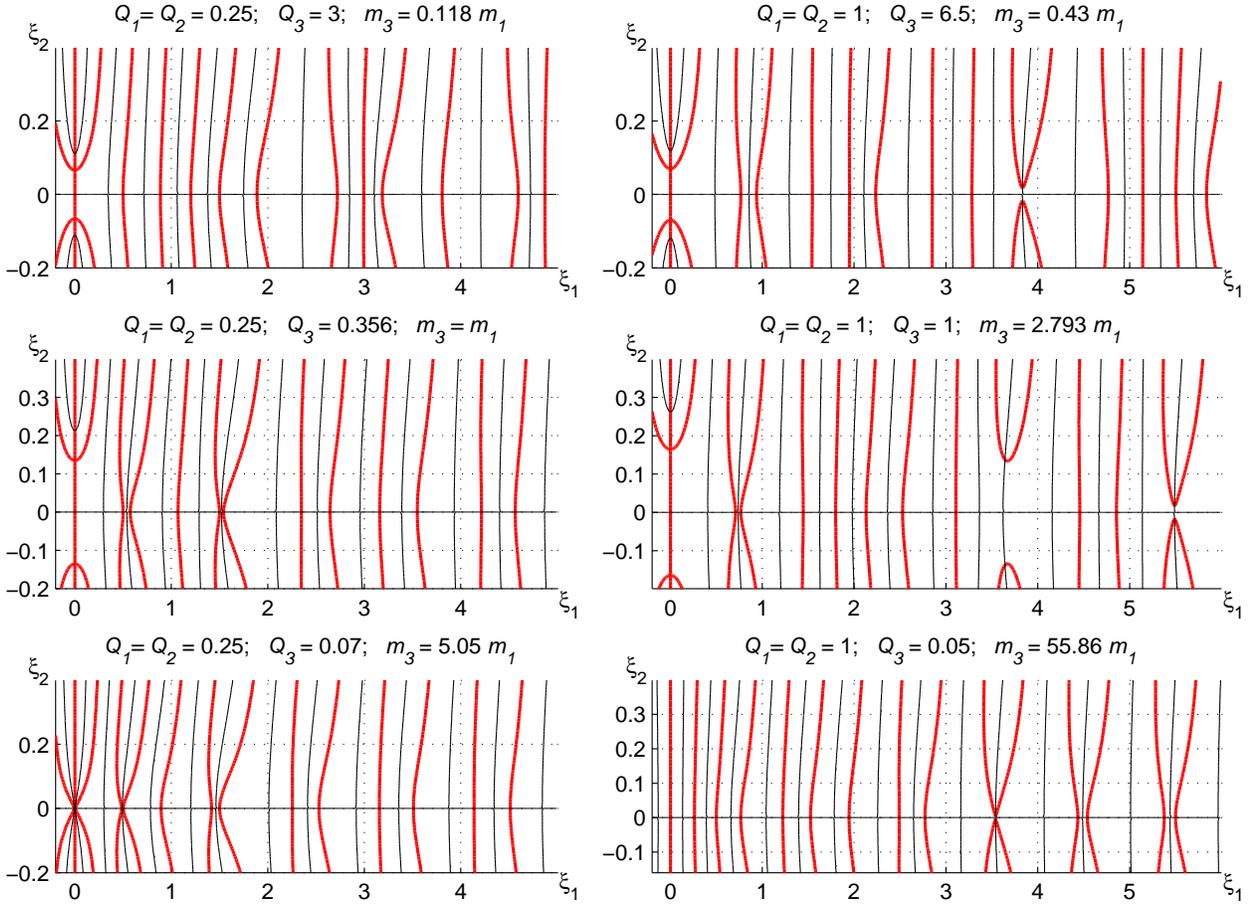}
\caption{Zero level lines for real part (thick) and imaginary part
(thin) of Eq.~(\ref{frmm2}) for specified values $Q_j$ and $m_3$}
\end{figure}

Fig.~2 demonstrates, that Eq.~(\ref{frmm2}) has complex roots
$\xi=\xi_1+i\xi_2$ with positive imaginary parts $\xi_2>0$, if the
value $Q_3$ (and the corresponding mass $m_3$) is not too small or
too large. These roots generate exponentially growing modes of
disturbances: $|\f|\sim\exp(\xi_2\tau)$.

The same picture takes place for other values $\om$. So we may
conclude, that the cental rotational states (\ref{Xlin}) of the
type (\ref{mm}) are {\sl unstable} with respect to small
disturbances, if the central mass is restricted by
 \be
 0<m_3<m_{3cr}.
 \label{m3cr}\ee
 The critical value $m_{3cr}$ (and the corresponding value
$Q_{3cr}$) is determined from the condition of vanishing all
complex roots of Eq.~(\ref{frmm2}) (with Im\,$\xi>0$) for
$m_3>m_{3cr}$. Thus, we obtain the threshold effect in stability
properties: if $m_3\ge m_{3cr}$, all roots of Eq.~(\ref{frmm2})
are real ones and the state is stable. But in the case
$m_3<m_{3cr}$ the state is unstable.

For any values $Q_1=Q_2$ and any (arbitrarily small) value $m_3$
from the interval (\ref{m3cr}) there are exists the pure imaginary
root $\xi^*=i\xi_2^*$ ($\xi_2^*>0$). It tends to 0 at $m_3\to0$.
For the states with $\om\le0{.}5$, $Q_1\le0{.}25$ equation
(\ref{frmm2}) has no other complex roots except $\xi^*=i\xi_2^*$.
If mass $m_3$ increases, the value $\xi_2^*$ (increment of
disturbances' growth) increases too, reaching the maximal value
for masses $m_3$, close to $m_1$ in order of magnitude. If $m_3$
increases further ($Q_3$ diminishes) the increment $\xi_2^*$
decreases and vanishes at the critical value $m_{3cr}\equiv
m_{3cr}^*=\gamma a_0/Q_{3cr}^*$,  in the (\ref{m3cr}). For the
case $Q_1=Q_2=1/4$ the value $m_{3cr}^*\simeq5{.}05\,m_1$.

The critical value $Q_{3cr}^*$, corresponding to vanishing the
root $\xi^*$, may be calculated, if we substitute $\xi=i\xi_2$
into Eq.~(\ref{frmm2}) and analyze its behavior at $\xi_2\to0$:
 $$
\bs_1Z_1\om^5\xi_2-2Q_3Z_1\om^4\frac{\bc_1(1+\cosh\pi\xi_2)+
\bs_1\om\sinh\pi\xi_2}{\cosh\pi\xi_2}+\phi(\xi_2)=0,\qquad\phi(\xi_2)=
{\cal O}(\xi_2^3).
 $$
 The function $\phi(\xi_2)=\phi_3\xi_2^3+\phi_5\xi_2^5+\dots$
is positive for $\xi_2>0$ (contains only positive summands), so
the root $\xi_2=\xi_2^*$ of this equation exists only under the
condition $2Q_3(2\bc_1+\pi \bs_1\om)>\bs_1\om$. From this
condition and equalities (\ref{h1h2}) and (\ref{mm}) one can find
the critical value $Q_{3cr}=Q_{3cr}^*$ for the root $i\xi_2^*$:
 \be
Q_{3cr}^*=\frac1{2\pi+4\bc_1(\om\bs_1)^{-1}}=
\frac1{2\pi+2Q_1^{-1}}.
 \label{Q3mm}\ee

In the limit $m_1\to0$, ($Q_1\to\infty$) expression (\ref{Q3mm}) takes the form,
coinciding with the critical value in the condition of instability
$Q>Q_{cr}^*=(2\pi)^{-1}$ \cite{clstab05} for the cental rotational states of the closed
string with $n=1$ massive point. The equation of the small disturbances spectrum for this
string \cite{clstab05}
 $$
\xi^3+2Q(2\xi^2+\om^2)\tan\pi\xi+4Q^2\xi\tan^2\pi\xi-\om^2\xi=0
 $$
 results from Eq.~(\ref{frmm2}) ($Q\equiv Q_3$) in the mentioned
limit $m_j\to0$, $j=1,2$.

For values $\om>1/2$ the structure of complex roots of Eq.~(\ref{frmm2}) is more
complicated. Because of the symmetry we shall count only roots in the quadrant
$\xi_1\ge0$, $\xi_2>0$. For $Q_1>1/4$ in certain interval of values $m_3$ the second
complex root of Eq.~(\ref{frmm2}) exists (denoted below by $\xi^\diamond$); for larger
values $Q_1$ corresponding to $v_1\to1$ the third and other roots appear. In particular,
in Fig.~2 for $Q_1=1$ the root $\xi^\diamond$ appears, if $Q_3<6{.}667$; for $Q_3=1$ the
complex roots are: $\xi^*\simeq0{.}262\,i$; $\xi^\diamond\simeq3{.}64+0{.}137\,i$; the
third root $\xi^\triangle\simeq5{.}49+0{.}017\,i$ (it exists in the narrow interval
$0{.}86<Q_3<1{.}65$).

For different values $Q_1$ the critical value $Q_{3cr}$ is
determined by vanishing the root $\xi^*$ or $\xi^\diamond$. For
the states in Fig.~2 for $Q_1=1/4$ the first variant takes place
and $Q_{3cr}=Q_{3cr}^*\simeq0{.}07$; for $Q_1=1$ we see the second
variant: $Q_{3cr}=Q_{3cr}^\diamond\simeq0{.}05$.

In the case $m_3=0$, corresponding to $Q_3\to\infty$, there is no
massive point at the center, and we have the linear rotational
state with $n=2$. In this case equation (\ref{frmm2}) takes the
form
 \be
\Big(\xi+\om\tan\frac{\pi\om}2\tan\frac{\pi\xi}2\Big)\xi\sin\pi\xi=0.
 \label{frmmo}\ee
 For all values $\om$ it has only real roots. So the
linear rotational state with $n=2$ of the type (\ref{mm}) are
stable.

Stability also takes place for the case $Q_3=0$ ($m_3\to\infty$).

Generalization of the above analysis for the case of arbitrary
masses, in particular, $m_1\ne m_2$ results in rather complicated
equation of the small disturbances spectrum. In the particular
case $Q_3=0$ ($m_3\to\infty$) this equation have the form
 \be\!\!\!
\begin{array}{c}
\bigg\{\tilde s\tilde s_1\tilde s_3\bc_1^3\bc_3^3 \xi^6 -3\tilde
s_1(\tilde s\tilde c_3+\tilde c\tilde s_3)
\bs_3\bc_1^3\bc_3^2\om\xi^5
 +\tilde s_1\Big[\tilde c\tilde c_3\bs_3(4\bc_1\bs_3+9\bs_1\bc_3)
-2\tilde s\tilde s_3\bc_1Z_3\Big]\bc_1^2\bc_3\om^2\xi^4+\\
 +\Big[\tilde s\tilde s_1\tilde c_3\bs_3(6\bc_1\bc_3\bs_1\bs_3+\bc_1^2Z_3+3\bc_3^2Z_1)
 +\tilde c\Big(\tilde s_1\tilde s_3Z_3(\bc_1\bs_3+3\bs_1\bc_3)-
 6\tilde c_1\tilde c_3\bc_3\bs_1\bs_3^2\Big)\bc_1\Big]\bc_1\om^3\xi^3-
 \rule{0mm}{1.2em}\\
 -\Big[\tilde c(3\tilde s\bc_1\bs_3+4\tilde c_1\tilde s_3\bs_1\bc_3)\,
 Z_3\bs_1+2\tilde s_1\tilde c_3(4\tilde c_1\tilde c_3\bs_1^2\bs_3^2-
 \tilde s_1\tilde s_3Z_1Z_3)\,\bc_3\Big]\bc_1\om^4\xi^2+\rule{0mm}{1.2em}\\
 +\Big[\tilde c(2\tilde c_1\tilde c_3\bs_1^2-\tilde s_1\tilde s_3Z_1)\,Z_3
 \bc_1\bs_3-\tilde s\tilde c_1\tilde s_3\bs_1Z_3(2\bc_1\bs_1\bs_3+\bc_3Z_1)\Big]
 \om^5\xi+\tilde c\tilde s_1\tilde c_3\bs_1\bs_3Z_1Z_3\om^6+\rule{0mm}{1.2em}\\
 +(1\leftrightarrow3)\bigg\}\xi(\xi^2-\om^2)=0.
\end{array}\!\!\!\!\!\!
  \label{frgeno}\ee
 Here
$\,Z_3=1+\bs_3^2$, $\,\tilde c_3=\cos(\pi-\s_1)\,\xi$, $\,\tilde
s_3=\sin(\pi-\s_1)\,\xi,\,$
 the symbol $(1\leftrightarrow3)$ means repeating the same terms
with transposed indices ``1'' and ``3''.

Equation (\ref{frgeno}) has only real roots, so the rotational
states (\ref{Xlin}) with the infinitely heavy mass $m_3\to\infty$
at the center are stable.

Linear rotational states with $n=2$ ($m_3=0$) are also stable for
any values $\om$, $m_1$, $m_2$.

But for intermediate values $m_3$ from the interval (\ref{m3cr})
the spectrum of small disturbances contains complex frequencies
$\xi=\xi_1+i\xi_2$. Their behavior is presented in Fig.~3.

\begin{figure}[hb]
\includegraphics[scale=0.85,trim=30 23 10 35]{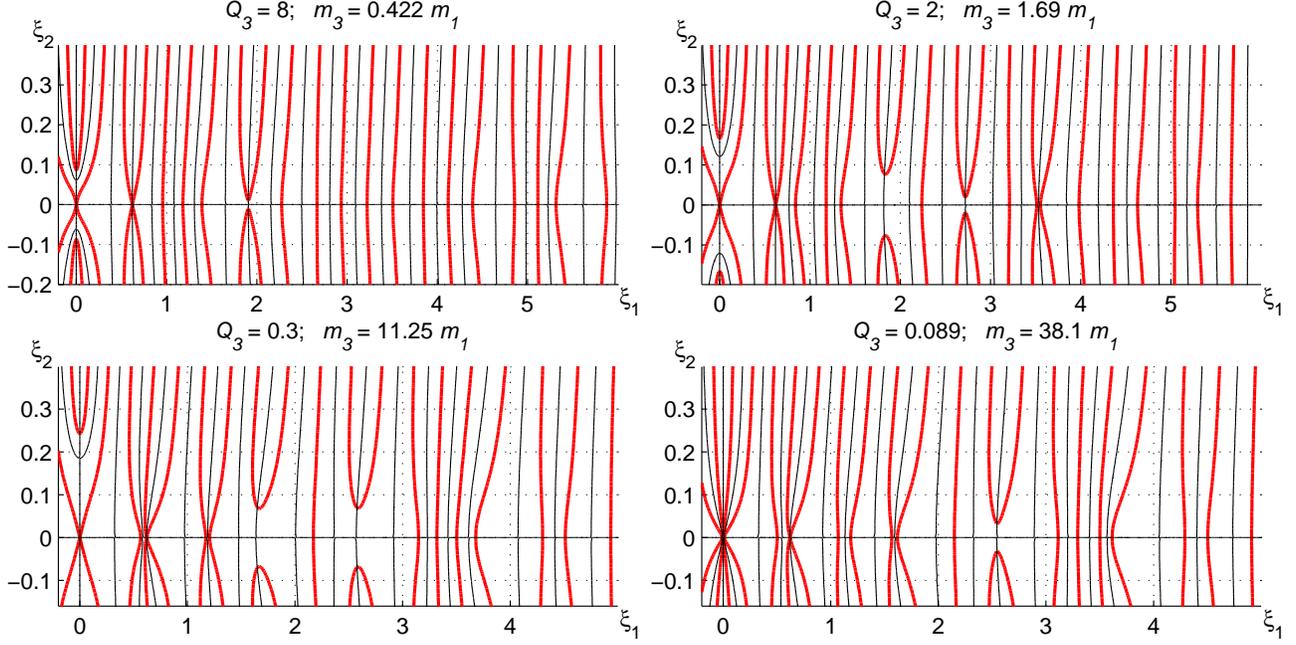}
\caption{Frequencies of small disturbances for the state with
$Q_1=1$, $Q_2=\frac14$ ($m_2\simeq10{.}5\,m_1$) and  specified
$Q_3$, $m_3$}
\end{figure}

For this state $\om\simeq0{.}62025$. Behavior of complex roots is
similar to the case (\ref{mm}): there is the range (\ref{m3cr})
$0<m_3<m_{3cr}$ of existence of complex roots, hence the
rotational state is unstable in this range.

There are three types of complex roots in Fig.~3 (similarly to
Fig.~2): the imaginary root $\xi^*$, the complex roots
$\xi^\diamond$ and $\xi^\triangle$. If we suppose $m_1=1$ (it is
equivalent to $\gamma a_0\simeq3{.}376$), that the range of
existence $\xi^*$ is $0<m_3<38{.}1$, for $\xi^\diamond$ it is
$0{.}42<m_3<30{.}7$, and for $\xi^\triangle$ this range is
$1{.}64<m_3<m_{3cr}\simeq67{.}5$. For other parameters of the
state the critical mass $m_{3cr}$ is determined from vanishing the
root $\xi^*$ or other roots.

Generalization of the expression (\ref{Q3mm}) for the critical
value $Q_{3cr}^*$, determined from vanishing the root $\xi^*$
takes the form
 $$
\big(Q_{3cr}^*\big)^{-1}=2\pi+Q_1^{-1}+Q_2^{-1}.
 $$
 Taking into account Eq.~(\ref{a0v}) $m_3=\gamma a_0/Q_3$ we
obtain  the critical value of the central mass
 \be
m_{3cr}^*=2\pi\gamma a_0+\frac{m_1}{\sqrt{1-v_1^2}}
+\frac{m_2}{\sqrt{1-v_2^2}}\equiv E-m_3.
 \label{m3crE}\ee
 It coincides with energy of this state of the string without contribution
of the mass $m_3$ \cite{Tr,clbgl07}.

We may conclude, the central rotational state is unstable if the
central mass $m_3$ is nonzero and less than energy of the string
with other massive points.

\bigskip

\noindent{\bf 5. Numerical experiments}

\medskip

In this section the above results connected with stability or
instability of rotational states are verified in numerical
experiments. Slightly disturbed rotational states (\ref{Xlin}) of
the closed string with $n=3$ masses are simulated numerically in
comparison with the similar states of the linear string baryon
model $q$-$q$-$q$ \cite{lin}. We use the approach, suggested in
Refs.~\cite{stab,stlin,Y02}. It includes solving the
initial-boundary value problem, by other words, calculation of
classical motion of the system on the base of given initial
position of the string $X^\mu\big|_{ini}=\rho^\mu(\tilde\s)$ and
initial velocities of its points $\dot
X^\mu\big|_{ini}=v^\mu(\tilde\s)$, $\tilde\s=\tilde\s(\s)$.

\begin{figure}[bh]
\includegraphics[scale=0.8, trim=6mm 0mm 0mm 2mm]{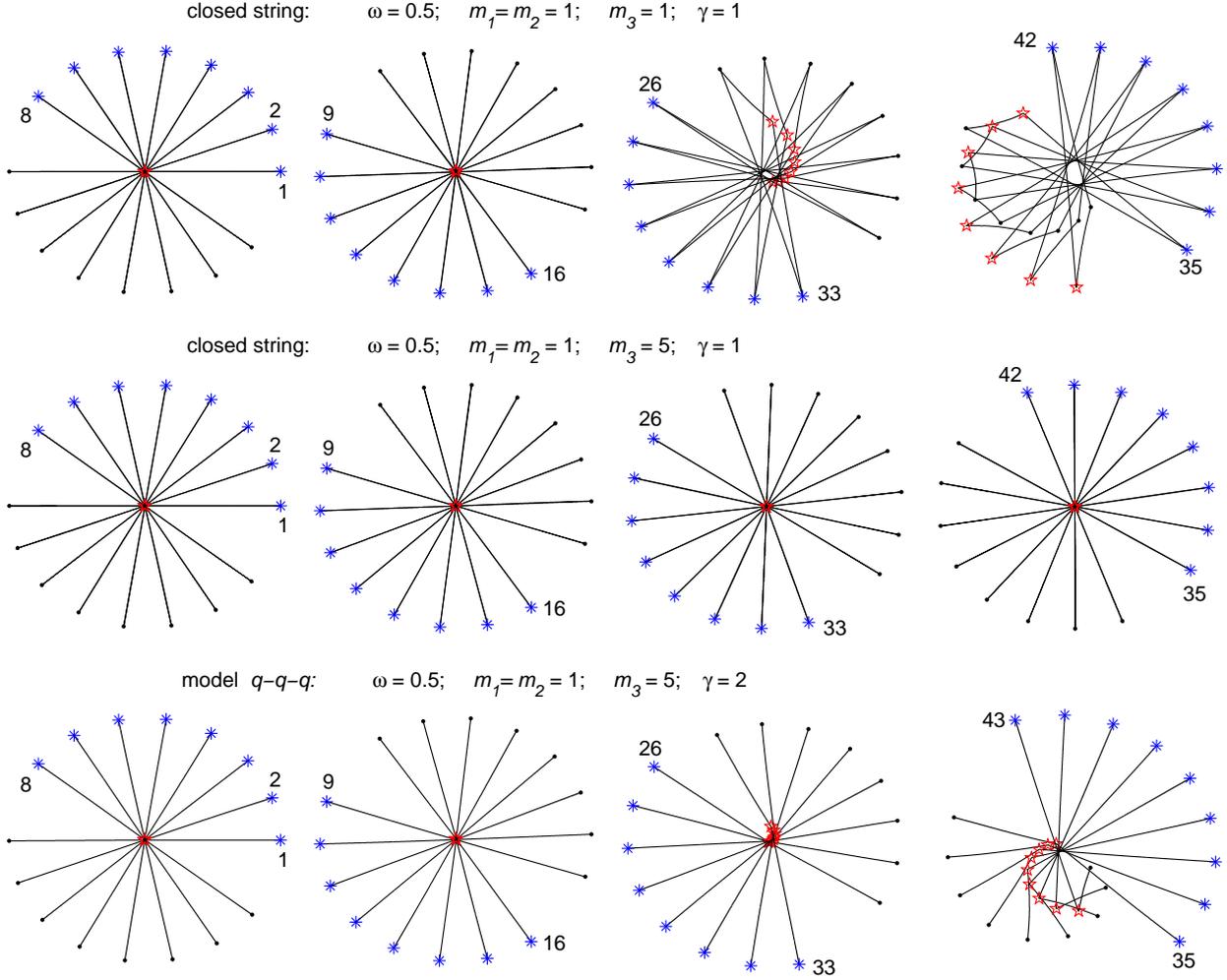}
\caption{Numerical simulation of disturbed rotational motions}
\end{figure}

If this initial conditions correspond to a central rotational
state (\ref{Xlin}) with a small disturbance, for example
$\;\rho^\mu(\tilde\s)=\rho^\mu_{rot}(\tilde\s)+\de\rho^\mu(\tilde\s)$,
$\;v^\mu(\tilde\s)=v^\mu_{rot}(\tilde\s)+\de v^\mu(\tilde\s)$,
 we calculate slightly disturbed rotational motion of the system.
In Fig.~5 the examples of such a motion for the closed string and
the linear model $q$-$q$-$q$ are represented as a set of
``photographs'' or positions of the string in $e_1,e_2$-plane.
These positions (sections $t={}$const of the world surface) are
numbered in order of increasing $t$ with spacing in time $\Delta
t=0.25$; the numbers are shown near the first massive point marked
by the asterisk. The second and the third (central) masses are
marked by the point and the pentagram correspondingly. Parameters
of the states are pointed out, they correspond to
$Q_1=Q_2=\frac14$ (compare with Fig.~2).

In three cases in Fig.~4 the small disturbance in initial data
 is in the form $\de\rho^\mu=0$ and $\de
v^\mu(\tilde\s)=0{.}01\sin(\tilde\s-\tilde\s_2)$ in the interval
$\tilde\s_2<\tilde\s<\tilde\s_3$ ($\tilde\s_3-\tilde\s_2=\pi$)
between the points $m_2$ and $m_3$, but $\de v^\mu$ equals zero on
all other points.

As one can see in Fig.~4, if the value $m_3$ is in the interval
(\ref{m3cr}), and the increment $\xi_2^*$ of disturbances' growth
is large enough (the case with $m_3=1$), the evolution of growing
disturbances results in going away the central mass. The string
changes into rotating curvilinear triangle and massive points
change their positions.

In the case $m_3\ge m_{3cr}$ the motion is stable and the massive
point $m_3$ remains near the rotational center. Similar picture
takes place, if $m_3\simeq m_{3cr}$, in particular $m_3=5$ is
close to $m_{3cr}\simeq5{.}05$. In this case the increment
$\xi_2^*$ is very small and the disturbed motion looks like a
stable one.

It is interesting to compare this picture with slightly disturbed
rotational states for the linear string baryon model $q$-$q$-$q$.
If tension $\gamma$ is twice large than for the closed string and
masses $m_j$ are the same, parameters of a central rotational
state $\om$, $Q_j$ and $v_j$ coincide for both models. It was
shown in Ref.~\cite{stablin} that the spectrum of small
disturbances (\ref{Psi+f}) for these states of the system
$q$-$q$-$q$ is described by the equation
 $$
\frac{m_3(1-v_i^2)\,\xi(\xi^2-\om^2)}{m_1v_1\om(\xi^2+\om^2)}=
\frac{\bc_1(Q_1^2\kappa_1-\xi^2)-2\bs_1Q_1\xi}
{\bs_1(Q_1^2\kappa_1-\xi^2)+2\bc_1Q_1\xi}+
\frac{\bc_3(Q_2^2\kappa_2-\xi^2)-2\bs_3Q_2\xi}
{\bs_3(Q_2^2\kappa_2-\xi^2)+2\bc_3Q_2\xi},
 $$
 where $\kappa_j=1+v_j^{-2}$. For any value $m_3>0$
the imaginary root $\xi=i\xi_2^*$ of this equation exists (there
is no critical maximal value), so the picture of stability differs
from that for the closed string: central rotational states of the
linear string baryon model $q$-$q$-$q$ are unstable for any masses
$m_3>0$.

\bigskip

\centerline {\bf 6. Regge trajectories for unstable states}
\medskip

Rotational states of the closed string with $n$ massive points are
applied for describing orbitally excited baryons \cite{4B,InSh}
and the Pomeron trajectory \cite{clbgl07,GlueY08}, corresponding
to possible glueball states.

For linear and central rotational states (\ref{Xlin}) the energy
$E$ and angular momentum $J$ are used in
Refs.~\cite{4B,clbgl07,GlueY08} in the following form:
 \be
E=2\pi\gamma  a_0+\sum_{j=1}^n \frac{m_j}{\sqrt{1-v_j^2}}+\Delta
E_{SL},
 \label{E}\ee
 \be
 J=L+S=
\frac{\gamma a_0^2}{2\omega}\bigg(2\pi+\sum_{j=1}^n
\frac{v_j^2}{Q_j} \bigg)+\sum_{j=1}^ns_j.
 \label{J}\ \ee
 Here $s_j$ are spin projections of massive points (quarks or valent gluons),
$\Delta E_{SL}$ is the spin-orbit contribution to the energy in
the following form \cite{4B}:
 $$
\Delta E_{SL}=\sum_{j=1}^n\big[1-(1-v_j^2)^{1/2}\big] (
\textbf{$\Omega$}\cdot \textbf{s}_j).
 $$

If the string tension $\gamma$, values $m_j$ and the type of
rotational state are fixed, we obtain the one-parameter set of
motions with different values $E$ and $J$. The parameter of this
set is any of the values: $\omega$, $a_0$, $E$, $J$; other values
are expressed via relations (\ref{s21})\,--\,(\ref{oms1}). These
states lay at quasilinear Regge trajectories. If a central (or
linear) rotational state (\ref{Xlin}) is the simplest one, that is
$k_1=k_2=0$ in Eq.~(\ref{oms1}) the asymptotic behavior of the
corresponding Regge trajectory  in the limit $E\to\infty$ is
\cite{clbgl07,GlueY08}:
 \be
J\simeq\al'(E-m_3)^2,\qquad \al'=\frac1{4\pi\gamma}.
 \label{JElim}\ee

Below we apply the central rotational states to describing the
Pomeron trajectory \cite{GlueY08,DonPom}, corresponding to
glueball states. We suppose that the value $S$ in Eq.~(\ref{J})
corresponds to the maximal total momentum (\ref{J}), that is $S=2$
for 2-gluon glueballs and $S=3$ for 3-gluon glueballs
\cite{AbreuB,MathSS3}. Other values of model parameters are
\cite{clbgl07,GlueY08}:
 \be
\gamma=0{.}175\mbox{ GeV}^2,\quad m_1=m_2=m_3=750\mbox{ MeV}.
 \label{gamm12}\ee
 This tension $\gamma$  corresponds to the slope of Regge
trajectories for hadrons $\alpha'\simeq0{.}9$ GeV$^{-2}$.
Estimations of gluon masses on the base of gluon propagator, in
particular, in lattice calculations \cite{BernardBS} yield values
$m_j$ from 700 to 1000 MeV.

\begin{figure}[bh]
\includegraphics[scale=0.8, trim=-33mm 14mm 0mm 26mm]{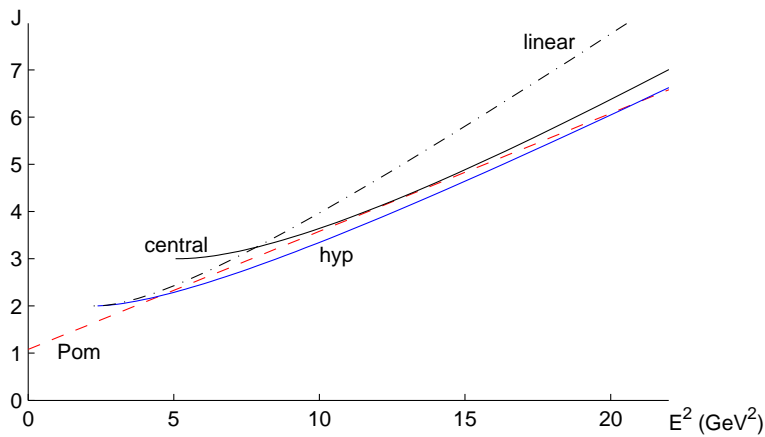}
\caption{ Regge trajectories for rotational states}
\end{figure}

Regge trajectories or graphs $J=J(E^2)$ of expressions (\ref{E})
and (\ref{J}) for linear, central and hypocycloidal rotational
states with parameters (\ref{gamm12}) are presented in Fig.~5.
These trajectories lie close to the pomeron trajectory
\cite{DonPom,MeyerT}
 \be
 J\simeq1{.}08+0{.}25E^2;
 \label{ReggPom} \ee
it is shown as the dashed line. The dash-dotted line corresponds
to the linear state with $n=2$, the hypocycloidal state is
``triangle'' type state with $n=2$ (the third vertex of the
triangle is the massless point), considered in
Refs.~\cite{clbgl07,GlueY08}. For the central rotational state
$n=3$.

These Regge trajectories are nonlinear for small $E$ and tend to
linear if $E\to\infty$. Their slope $\al'$ in this limit tends to
the limit (\ref{JElim}) for linear and central rotational states
and to the value $\al'=\frac38(2\pi\gamma)^{-1}$ for ``triangle''
hypocycloidal states \cite{GlueY08}.

We have shown in Sect.~4 that the central rotational states
(\ref{Xlin}) with equal masses (\ref{gamm12}) are unstable for all
energies on the classic level. But this does not mean
disappearance or terminating corresponding Regge trajectories in
Fig.~5. The straight consequence of this instability is the
contribution to width of a hadron state.

String models describe only excited hadron states with large
orbital momenta $L$. These states are unstable with respect to
strong interactions and have rather large width $\Gamma$. In
string interpretation of excited hadron this width is connected
with probability of string breaking; this probability is
proportional to the string length $\ell$ \cite{Kodecay,GuptaR}.
The value $\ell$ is proportional to the string contribution
$E_{str}$ to energy $E$ of a hadron state, in particular, for
rotational states (\ref{Xlin}) this contribution to the expression
(\ref{E}) is $E_{str}=2\pi\gamma  a_0$.

Therefore, the component of width $\Gamma_{br}$, connected with
string breaking, is proportional to $E_{str}$ with the factor
$0{.}1$ resulting from particle data \cite{GuptaR,KlemptZ}:
 \be
 \Gamma_{br}\simeq0{.}1\cdot E_{str}=0{.}2\cdot\pi\gamma  a_0.
 \label{Gambr} \ee

The contribution  $\Gamma_{inst}$ to width $\Gamma$ due to
instability of  string central rotational states (\ref{Xlin}) with
respect to small disturbances is determined from the increment
$\xi_2$ of exponential growth
 $$
 |\f|\sim\exp(\xi_2\tau)=\exp(\xi_2a_0^{-1}t).
 $$
 For the central state in Fig.~5 with parameters (\ref{gamm12})
 the increment $\xi_2=\xi_2^*$, so its width is
 \be
 \Gamma_{inst}\simeq\frac{\xi_2^*}{a_0}.
 \label{Gaminst} \ee

The values $\xi_2^*$ and $\Gamma_{inst}$ are calculated for the
mentioned central state in Fig.~5 for various energies  and the
graph $\Gamma_{inst}=\Gamma_{inst}(E)$ is presented in Fig.~6{\it
a} as the dashed line (here $E$ is energy without spin-orbit
correction). One can compare it with the value $\Gamma_{br}(E)$
(\ref{Gambr}), connected with string breaking (dash-dotted line),
and with the total width (solid line):
 \be
 \Gamma(E)=\Gamma_{br}(E)+\Gamma_{inst}(E).
 \label{Gam} \ee

\begin{figure}[bh]
\includegraphics[scale=0.8, trim=5mm 5mm 0mm 53mm]{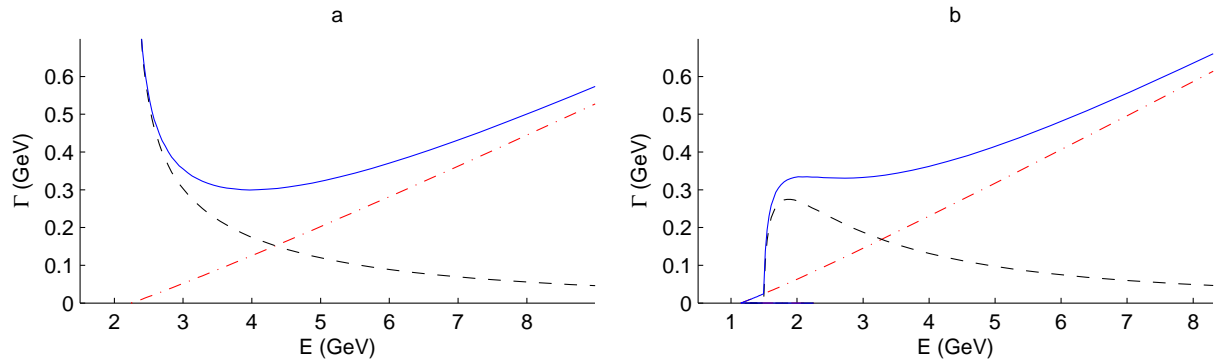}
\caption{Width $\Gamma(E)$ (\ref{Gam}) (solid line) as the sum of
$\Gamma_{br}$ (\ref{Gambr}) (dash-dotted line) and $\Gamma_{inst}$
(\ref{Gaminst}) (dashed line) for central states (a) with
parameters (\ref{gamm12}); (b) with $m_1=m_2=200$ MeV}
\end{figure}

In Fig.~6{\it a} we see that the central state with equal masses
(\ref{gamm12}) is unstable for all energies $E$. The corresponding
width component  $\Gamma_{inst}(E)$ (\ref{Gaminst}) tends to zero
at large $E$ (because $a_0$ increases) and tends to infinity in
the limit $E\to E_{min}=\sum m_j$. In this limit both values
$\xi_2^*\to0$, $a_0\to0$ but their ratio $\Gamma_{inst}\to\infty$.
So in the limit $E\to E_{min}$ the central states with
$m_3<m_1+m_2$ are unstable (note that string models are not
applicable in this limit).

Behavior of width $\Gamma(E)$ (\ref{Gam}) for central rotational
states of the system with $m_3>m_1+m_2$ is presented in Fig.~6{\it
b} with the same notations. Here $m_3$ and $\gamma$ are the same
values (\ref{gamm12}). In this case the threshold effect
(\ref{m3crE}) of instability exists, so the ``instability'' width
$\Gamma_{inst}(E)$ equals zero for energies $E<E_{cr}=2m_3$ (here
it is 1{.}5 GeV). For $E>E_{cr}$ the value $\Gamma_{inst}(E)$
exceeds $\Gamma_{br}(E)$ in the certain interval, but if $E$
increases, $\Gamma_{inst}(E)$ tends to zero and $\Gamma_{br}(E)$
increases linearly.

\bigskip

\centerline {\bf Conclusion}
\medskip

For linear rotational states (\ref{Xlin}) of a closed string with 2 massive points and
central rotational states (with the mass $m_3$ at the rotational center) the stability
problem is solved on the classic level. It is shown that the linear states are stable
with respect to small disturbances in linear approximation, but the central states are
unstable, if the central mass it less than the critical value (\ref{m3crE}). This value
equals energy of the string with other massive points.

This threshold effect was investigated both analytically
(instability is connected with exponentially growing modes in the
spectrum of small disturbances of a rotational state) and in
numerical experiments.

Instability of these central states results in some manifestations, in particular, in
additional width of excited hadron states. It is shown that this contribution makes
essential changes in linear dependence $\Gamma\sim E$.

\medskip

\centerline{\bf Acknowledgment}

\medskip

The author is grateful to Russian foundation of basic research for
the support (grant 05-02-16722).

\end{document}